\documentclass[aps,prd,twocolumn,superscriptaddress,showpacs]{revtex4}

\usepackage{graphicx}% Include figure files
\usepackage{dcolumn}% Align table columns on decimal point
\usepackage{setspace}
\usepackage{subfigure}
\usepackage{slashbox}

\begin{document}

\newcommand{\ttbar}{\mbox{$t\bar{t}$}}
\newcommand{\thetastar}{\mbox{$\theta^*$}}
\newcommand{\costhetastar}{\mbox{$\cos\thetastar$}}
\newcommand{\Mlb}{\mbox{$M_{lb}$}}
\newcommand{\Et}{\mbox{$E_T$}}
\newcommand{\Pt}{\mbox{$p_T$}}
\newcommand{\Flong}{\mbox{$F_0$}}
\newcommand{\Fplus}{\mbox{$F_+$}}
\newcommand{\met}{\mbox{$\protect \raisebox{0.3ex}{$\not$}\Et$}}

% You should use BibTeX and revtex.bst for references
\bibliographystyle{revtex}

\vspace*{1.5cm}

\title{Measurement of the $t\bar t$ Production Cross Section in $p\bar p$ collisions at $\sqrt{s}=1.96$\,TeV in the All Hadronic Decay Mode}
\affiliation{Institute of Physics, Academia Sinica, Taipei, Taiwan 11529, Republic of China} 
\affiliation{Argonne National Laboratory, Argonne, Illinois 60439} 
\affiliation{Institut de Fisica d'Altes Energies, Universitat Autonoma de Barcelona, E-08193, Bellaterra (Barcelona), Spain} 
\affiliation{Baylor University, Waco, Texas  76798} 
\affiliation{Istituto Nazionale di Fisica Nucleare, University of Bologna, I-40127 Bologna, Italy} 
\affiliation{Brandeis University, Waltham, Massachusetts 02254} 
\affiliation{University of California, Davis, Davis, California  95616} 
\affiliation{University of California, Los Angeles, Los Angeles, California  90024} 
\affiliation{University of California, San Diego, La Jolla, California  92093} 
\affiliation{University of California, Santa Barbara, Santa Barbara, California 93106} 
\affiliation{Instituto de Fisica de Cantabria, CSIC-University of Cantabria, 39005 Santander, Spain} 
\affiliation{Carnegie Mellon University, Pittsburgh, PA  15213} 
\affiliation{Enrico Fermi Institute, University of Chicago, Chicago, Illinois 60637} 
\affiliation{Joint Institute for Nuclear Research, RU-141980 Dubna, Russia} 
\affiliation{Duke University, Durham, North Carolina  27708} 
\affiliation{Fermi National Accelerator Laboratory, Batavia, Illinois 60510} 
\affiliation{University of Florida, Gainesville, Florida  32611} 
\affiliation{Laboratori Nazionali di Frascati, Istituto Nazionale di Fisica Nucleare, I-00044 Frascati, Italy} 
\affiliation{University of Geneva, CH-1211 Geneva 4, Switzerland} 
\affiliation{Glasgow University, Glasgow G12 8QQ, United Kingdom} 
\affiliation{Harvard University, Cambridge, Massachusetts 02138} 
\affiliation{Division of High Energy Physics, Department of Physics, University of Helsinki and Helsinki Institute of Physics, FIN-00014, Helsinki, Finland} 
\affiliation{University of Illinois, Urbana, Illinois 61801} 
\affiliation{The Johns Hopkins University, Baltimore, Maryland 21218} 
\affiliation{Institut f\"{u}r Experimentelle Kernphysik, Universit\"{a}t Karlsruhe, 76128 Karlsruhe, Germany} 
\affiliation{High Energy Accelerator Research Organization (KEK), Tsukuba, Ibaraki 305, Japan} 
\affiliation{Center for High Energy Physics: Kyungpook National University, Taegu 702-701, Korea; Seoul National University, Seoul 151-742, Korea; and SungKyunKwan University, Suwon 440-746, Korea} 
\affiliation{Ernest Orlando Lawrence Berkeley National Laboratory, Berkeley, California 94720} 
\affiliation{University of Liverpool, Liverpool L69 7ZE, United Kingdom} 
\affiliation{University College London, London WC1E 6BT, United Kingdom} 
\affiliation{Centro de Investigaciones Energeticas Medioambientales y Tecnologicas, E-28040 Madrid, Spain} 
\affiliation{Massachusetts Institute of Technology, Cambridge, Massachusetts  02139} 
\affiliation{Institute of Particle Physics: McGill University, Montr\'{e}al, Canada H3A~2T8; and University of Toronto, Toronto, Canada M5S~1A7} 
\affiliation{University of Michigan, Ann Arbor, Michigan 48109} 
\affiliation{Michigan State University, East Lansing, Michigan  48824} 
\affiliation{Institution for Theoretical and Experimental Physics, ITEP, Moscow 117259, Russia} 
\affiliation{University of New Mexico, Albuquerque, New Mexico 87131} 
\affiliation{Northwestern University, Evanston, Illinois  60208} 
\affiliation{The Ohio State University, Columbus, Ohio  43210} 
\affiliation{Okayama University, Okayama 700-8530, Japan} 
\affiliation{Osaka City University, Osaka 588, Japan} 
\affiliation{University of Oxford, Oxford OX1 3RH, United Kingdom} 
\affiliation{University of Padova, Istituto Nazionale di Fisica Nucleare, Sezione di Padova-Trento, I-35131 Padova, Italy} 
\affiliation{LPNHE, Universite Pierre et Marie Curie/IN2P3-CNRS, UMR7585, Paris, F-75252 France} 
\affiliation{University of Pennsylvania, Philadelphia, Pennsylvania 19104} 
\affiliation{Istituto Nazionale di Fisica Nucleare Pisa, Universities of Pisa, Siena and Scuola Normale Superiore, I-56127 Pisa, Italy} 
\affiliation{University of Pittsburgh, Pittsburgh, Pennsylvania 15260} 
\affiliation{Purdue University, West Lafayette, Indiana 47907} 
\affiliation{University of Rochester, Rochester, New York 14627} 
\affiliation{The Rockefeller University, New York, New York 10021} 
\affiliation{Istituto Nazionale di Fisica Nucleare, Sezione di Roma 1, University of Rome ``La Sapienza," I-00185 Roma, Italy} 
\affiliation{Rutgers University, Piscataway, New Jersey 08855} 
\affiliation{Texas A\&M University, College Station, Texas 77843} 
\affiliation{Istituto Nazionale di Fisica Nucleare, University of Trieste/\ Udine, Italy} 
\affiliation{University of Tsukuba, Tsukuba, Ibaraki 305, Japan} 
\affiliation{Tufts University, Medford, Massachusetts 02155} 
\affiliation{Waseda University, Tokyo 169, Japan} 
\affiliation{Wayne State University, Detroit, Michigan  48201} 
\affiliation{University of Wisconsin, Madison, Wisconsin 53706} 
\affiliation{Yale University, New Haven, Connecticut 06520} 
\author{A.~Abulencia}
\affiliation{University of Illinois, Urbana, Illinois 61801}
\author{J.~Adelman}
\affiliation{Enrico Fermi Institute, University of Chicago, Chicago, Illinois 60637}
\author{T.~Affolder}
\affiliation{University of California, Santa Barbara, Santa Barbara, California 93106}
\author{T.~Akimoto}
\affiliation{University of Tsukuba, Tsukuba, Ibaraki 305, Japan}
\author{M.G.~Albrow}
\affiliation{Fermi National Accelerator Laboratory, Batavia, Illinois 60510}
\author{D.~Ambrose}
\affiliation{Fermi National Accelerator Laboratory, Batavia, Illinois 60510}
\author{S.~Amerio}
\affiliation{University of Padova, Istituto Nazionale di Fisica Nucleare, Sezione di Padova-Trento, I-35131 Padova, Italy}
\author{D.~Amidei}
\affiliation{University of Michigan, Ann Arbor, Michigan 48109}
\author{A.~Anastassov}
\affiliation{Rutgers University, Piscataway, New Jersey 08855}
\author{K.~Anikeev}
\affiliation{Fermi National Accelerator Laboratory, Batavia, Illinois 60510}
% ...
\collaboration{CDF Collaboration}
\noaffiliation

\begin{abstract}
We report a measurement of the $t\bar t$ production cross section using the CDF~II detector at the Fermilab Tevatron. The analysis is performed using 311~pb$^{-1}$ of $p\bar p$ collisions at $\sqrt{s}=1.96$~TeV. The data consist of events selected with six or more hadronic jets with additional kinematic requirements. At least one of these jets must be identified as a $b$-quark jet by the reconstruction of  a secondary vertex. 
The cross section is measured to be 
$\sigma_{\ttbar}=7.5\pm 2.1(\textnormal{stat.})^{+3.3}_{-2.2}(\textnormal{syst.})^{+0.5}_{-0.4}(\textnormal{lumi.})$\,pb,
which is consistent with the standard model prediction. 
\end{abstract}

\maketitle

\section{\label{sec:Intro}Introduction}
At the Tevatron, the dominant standard model mechanism for top quark production in $p\bar p$ collisions is predicted to be $q\bar q$ annihilation to $\ttbar$. The top quark decays immediately into a $W$ boson and a $b$ quark almost 100\% of the time. The $W$ boson subsequently decays to either a pair of quarks or a lepton-neutrino pair. 
The measurement of the $\ttbar$ cross section tests the QCD calculations for the pair production of a massive color triplet. These calculations have been performed in perturbation theory at the next-to-leading order\,\cite{topxs}. Recent work on corrections for soft gluon emission show that their effect on the cross section is small, and that they reduce the theoretical uncertainty due to the choice of renormalization and factorization scale. The total theoretical uncertainty is approximately 15\%. 
At $\sqrt{s} = 1.96$\,TeV, the predicted $\ttbar$ production cross section is 6.1\,pb for a top mass of  178~GeV/$c^2$, the average value of the Run I measurements\,\cite{RunIavemass}. 

In this analysis, we examine events with an all-hadronic final state characterized by a six-jet topology. In the standard model top decay, this final state has the advantage of a large branching ratio of $4/9$ and of being fully reconstructed. The major drawback is that it competes against a very large QCD multijet background which dominates the signal by three orders of magnitude after the application of the online trigger selection. 
To improve the signal-to-background ratio, a set of requirements based on the kinematic and topological characteristics of standard model $\ttbar$ events is applied to the data. In order to extract the $\ttbar$ signal, we select those jets identified as originating from $b$ quarks using a secondary vertex $b$-tagging algorithm, thus reaching a signal-to-background ratio of about 1/5. 
The CDF and D\O\, collaborations previously measured the $\ttbar$ production cross section %and top mass 
in the all-hadronic channel\,\cite{tophadPRL} using datasets with integrated luminosities of approximately 110\,pb$^{-1}$ collected at $\sqrt{s}=1.8$\,TeV during Run I of the Fermilab Tevatron Collider.
The results reported here are based on the data taken with the CDF II detector between March 2002 and September 2004, corresponding to an integrated luminosity of $311$~pb$^{-1}$. This measurement complements other recent $t\bar t$ cross section determinations by CDF in Run II using dilepton\,\cite{dilPRL} and lepton-plus-jets events\,\cite{ljetsPRD,ljetsKIN,ljetsBKIN,ljetsSLT}. The organization of the paper is as follows. Section\,\ref{sec:detector} contains a brief description of the CDF\,II detector. The trigger and the sample selections are described in Section\,\ref{sec:dataset} along with the acceptance associated with the optimized kinematic selection. The $b$-tagging algorithm and its efficiency for tagging $b$ jets are described in Section\,\ref{sec:btag}. In Section\,\ref{sec:bkg} the method for estimating the background from multijet processes is applied to the data and the related systematic uncertainties are evaluated. The $\ttbar$ production cross section measured in events with at least one $b$-tagged jet after the kinematic selection is presented in Section\,\ref{sec:xs} and the final result is summarized in Section\,\ref{sec:fine}.

\section{\label{sec:detector}The CDF II Detector}

The CDF\,II detector\,\cite{CDFdetector} is an azimuthally and forward-backward symmetric apparatus designed to study $p\bar p$ collisions at the Fermilab Tevatron. It uses a cylindrical coordinate system as described in\,\cite{coordinate}.
 It consists of a magnetic spectrometer surrounded by calorimeters and muon detectors. The charged particle tracking system is immersed in a 1.4~T magnetic field parallel to the $p$ and $\bar p$ beams. A set of silicon microstrip detectors provide charged particle tracking in the radial range from 1.5 to 28~cm. 
%%%
A 3.1~m long open-cell drift chamber, the central outer tracker (COT), covers the radial range from 40 to 137 cm. The COT provides up to 96 measurements of the track position with alternating axial and $\pm$2$^{\circ}$-stereo superlayers of 12-wire layers each. The fiducial region of the silicon detector extends to pseudorapidity $|\eta| \le 2$, while the COT provides coverage for $|\eta|\le 1$. Segmented electromagnetic and hadronic calorimeters surround the tracking system and measure the energy of interacting particles. The electromagnetic and hadronic calorimeters are lead-scintillator and iron-scintillator sampling devices, respectively, covering the range $|\eta|\le 3.6$. They are segmented in the central region ($|\eta|<1.1$) in towers of 15$^\circ$ in azimuth and 0.1 in $\eta$, and the forward region ($1.1<|\eta|<3.6$) in towers of 7.5$^\circ$ for $|\eta|<2.11$ and 15$^\circ$ for $|\eta|>2.11$.  
The electromagnetic calorimeters\,\cite{ecal,pem} are instrumented with proportional and scintillating strip detectors that measure the transverse profile of electromagnetic showers at a depth corresponding to the expected shower maximum.
The measured energy resolution for electrons in the electromagnetic calorimeters are $14\%/\sqrt{E_T}$\, in the central and $16\%/\sqrt{E_T}\oplus 1\%$\,in the forward\cite{TDR} where the units of $E_T$ are GeV.
We also measure the single-particle (pion) energy resolution in the hadronic calorimeters to be $75\%/\sqrt{E_T}$ for the central and $80\%\sqrt{E_T}\oplus 5\%$\, for the forward detector\,\cite{cha}.
Jets are identified as a group of electromagnetic and hadronic calorimeter clusters which fall within a cone of radius $\Delta R = \sqrt{\Delta\phi^2 + \Delta \eta^2}\le 0.4$\,\cite{jets}.
Drift chambers located outside the central hadronic calorimeters and behind a 60~cm iron shield detect muons with $|\eta| \le 0.6$\,\cite{CMU}. Additional drift chambers and scintillation  counters detect muons in the region $0.6<|\eta|<1.5$. Gas Cherenkov counters with a coverage of $3.7<|\eta|<4.7$ measure the average number of inelastic $p\bar p$ collisions and thereby determine the luminosity\,\cite{WZprl}. 
 
\section{\label{sec:dataset}Multijet Event Selection}

The all-hadronic final state of $\ttbar$ events is characterized by the presence of at least six-hadronic jets from the decay of the two top quarks. 
A multijet trigger relying on calorimetric information was specially developed to collect the events used in this analysis. 
After a preliminary selection of well contained and well reconstructed multijet events, tight kinematic requirements are imposed to reach a reasonable signal-to-background ratio. 

\subsection{Multijet Trigger Levels}

CDF uses a three-level trigger system, the first two consisting of special purpose electronics and the third level consisting of conventional computer processors.
For triggering purposes the calorimeter granularity is simplified to a $24\times 24$ grid in $\eta,\phi$ space and each trigger tower spans approximately 15$^\circ$ in $\phi$ and 0.2 in $\eta$ covering one or two physical towers.  
At level 1, a single trigger tower is required with $\Et\ge 10$~GeV, while at level 2 we require that the total transverse energy, summed over all the trigger towers, $\sum\Et$ be $\ge 125$~GeV and the presence of four or more clusters each with transverse energy $\Et\ge 15$~GeV. 
Finally, the third trigger level confirms the level 2 selection using more accurate determination of the jet energy, requiring four or more reconstructed jets with $\Et\ge 10$~GeV. %(with respect to $z=0$). 
This trigger rate corresponds to an effective cross section of about 14~nb and an efficiency of about 63\% for all $\ttbar$ events, and of about 85\% in the case of all-hadronic $\ttbar$ decays. 
The signal-to-background ratio (S/B) for $\ttbar$ events after this selection is about 1/3500 (assuming $\sigma_{\ttbar}=6.1$~pb).

\subsection{Preselection Requirements}

After full-event reconstruction, we retain only those events that are well contained in the detector.
We require the primary event vertex\,\cite{ljetsPRD} to be well reconstructed and to lie inside the luminous region ($|z|<60$~cm).
Jets are identified using a fixed-cone algorithm with a cone radius of 0.4 in $\eta-\phi$ space. 
The jet energies are corrected\,\cite{JESNIM} for detector response and multiple interactions. 
First, we take into account detector response variations in $\eta$, detector stability, and energy loss in the uninstrumented regions.
After a small correction for the extra energy deposited by multiple collisions in the same accelerator bunch crossing, a correction for calorimeter non-linearity is applied so that the jet energies correspond to the most probable in-cone hadronic energy. 
Each of these steps has an individual systematic uncertainty that is added in quadrature to derive the total uncertainty which decreases from 8 to 3\% with increasing jet energy. 
After these corrections the jet energy provides a good estimate of the initial parton energy. This can be verified comparing the jet energy to the energy of an electromagnetic object such as a prompt photon or a $Z$ boson produced in the same event. 
For this analysis, jets are required to have $\Et\ge 15$~GeV and $|\eta|\le 2$ after all corrections have been applied. 
We define the signal region by selecting events with a number of jets $6\le N_{\rm jets}\le 8$ in order to optimize the signal fraction.
In order to minimize the contamination of this sample from the $\ttbar$ leptonic channels, we veto events containing any well identified high-$\Pt$ electrons and muons as defined in\,\cite{ljetsKIN} and require that 
$\frac{\met}{\sqrt{\sum\Et}}$ be $<3$\,$\sqrt{{\rm GeV}}$\,\cite{metjetPRL}, 
where the  $\met$\,\cite{met} is corrected for both the momentum of any identified muons and the position of the $p\bar p$ collision point and the $\sum\Et$ is obtained by summing the $E_T$'s of all the selected jets. After these requirements 364,006 events are selected for further analysis. 

\subsection{Kinematic Selection Optimization and Acceptance}

\begin{figure}[htbp]
\centering
\subfigure{
\includegraphics[width=2.5in]{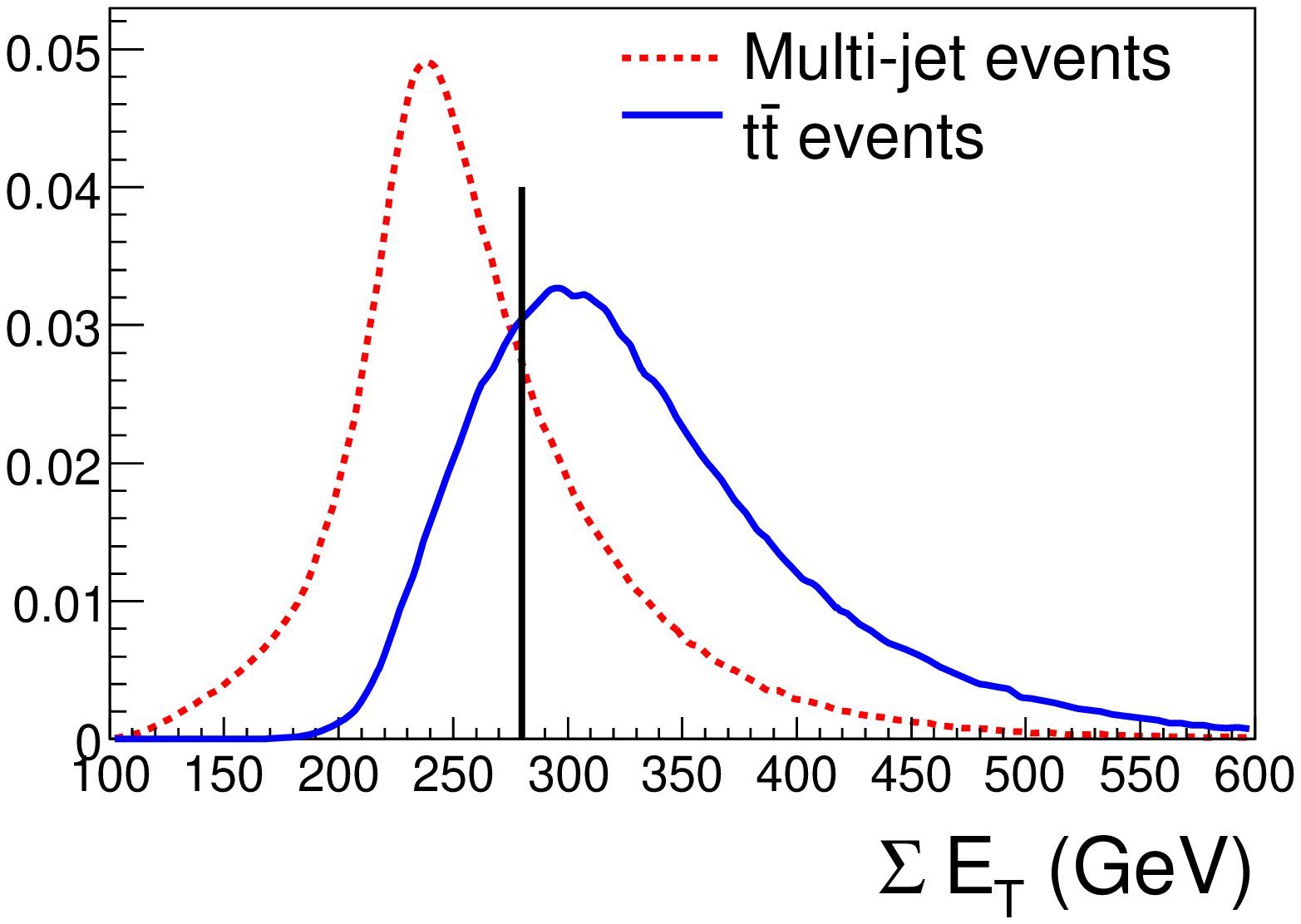}
}
%\hspace{4pt}
\subfigure{
\includegraphics[width=2.5in]{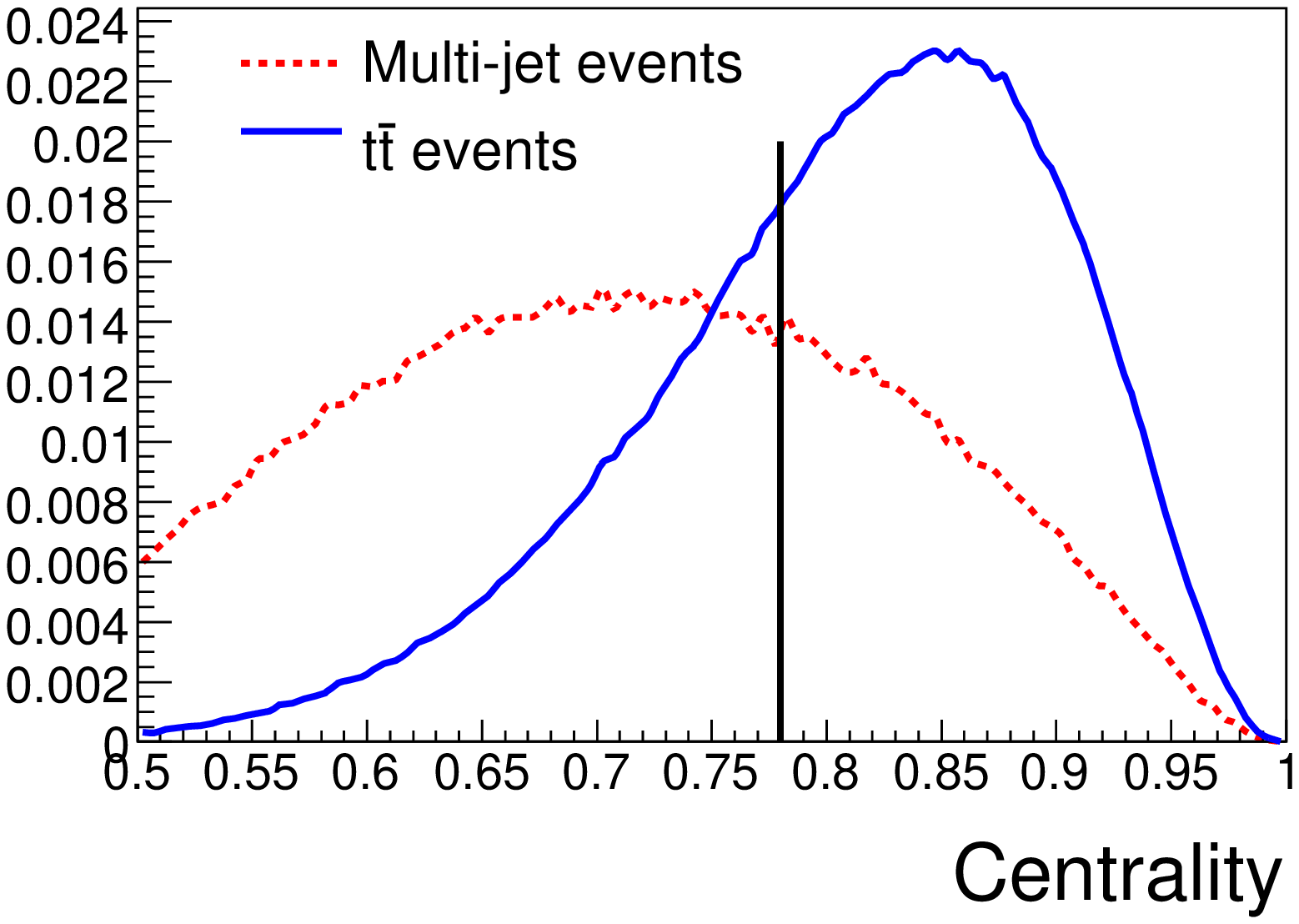}
}
%//
\subfigure{
\includegraphics[width=2.5in]{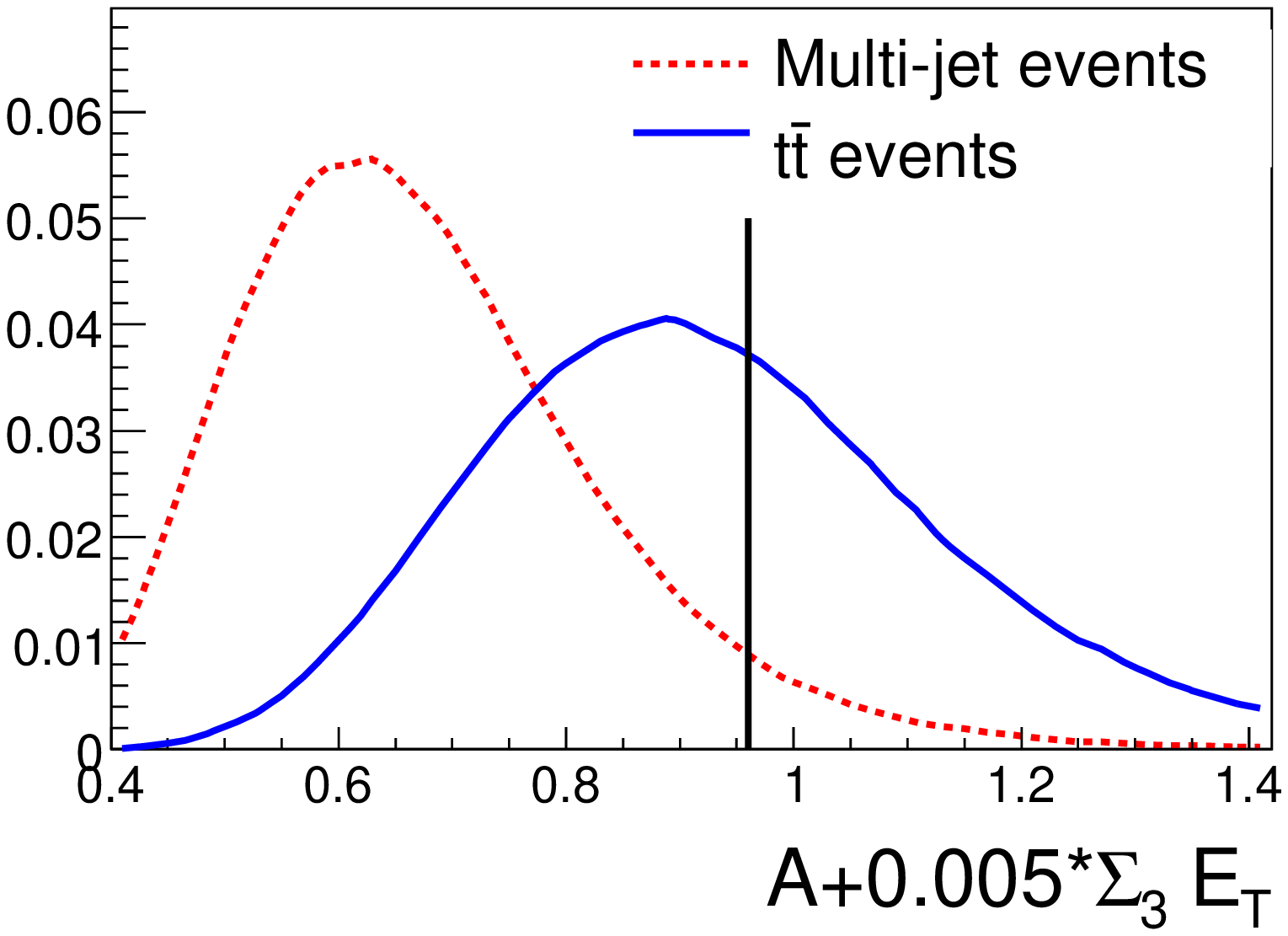}
}

\caption{Kinematic variable distributions in multijet data and $\ttbar$ Monte Carlo simulation. From top: $\sum E_T$, centrality, and A+0.005$\times \sum_3 E_T$. The lines represent the optimized selection cut. All histograms are normalized to unity.} \label{fig:kinsel1}
\end{figure}

We define a kinematic selection based on dynamical and topological properties of the event.
Quantities used are the number of jets, $N_{\rm jets}$, the total transverse energy of the jets, $\sum \Et$, and the quantity $\sum _3\Et\equiv \sum \Et - \Et^1 - \Et^2$, obtained by removing the contribution of the two jets with the highest $\Et$ from the total $\sum E_T$.
Other discrimimant variables considered are the centrality $C$, defined as $C = \frac{\sum E_T}{\sqrt{\hat{s}}}$, 
where $\sqrt{\hat{s}}$ is related to the energy of the hard scattering process as inferred from the all-jets invariant mass; and the aplanarity $A$, defined as ${A}=\frac{3}{2}{\cal{Q}}_1$, ${\cal{Q}}_1$ being the smallest of the three normalized eigenvalues of the sphericity tensor $M_{ab}= \sum _j P_{ja}P_{jb}$ calculated in the centre-of-mass system of all jets, where $P_j$ is the jet momentum.
In order to model the signal we use  {\sc PYTHIA}\,v6.2\,\cite{Pythia} and {\sc HERWIG}\,v6.4\,\cite{Herwig} leading-order Monte Carlo generators with parton showering followed by a simulation of the CDF II detector. The reference top mass chosen for the optimization is $m_{\rm top}=178$\,GeV/$c^2$. The background behavior is obtained from the multijet data events: this is possible since the signal fraction at the initial stage is very small, $\sim 4\%$ at most. 
Comparisons of the background-dominated data and Monte Carlo generated signal events for the chosen kinematic variables are shown in Fig.~\ref{fig:kinsel1}.

The kinematic selection is optimized for the maximum signal significance for $\ttbar$ events, defined as the ratio between the expected signal and the statistical uncertainty on the sum of signal and background. 
The values for the cuts after optimization are: ${\cal{A}}+0.005\sum _3\Et\ge 0.96$, $C\ge 0.78$, and 
$\sum \Et\ge 280$~GeV. 
Such a selection yields 3315 candidate events in the data with an efficiency of $6.7\pm 1.4\%$
for the $\ttbar$ signal and with a signal-to-background ratio $S/B\sim 1/25$. 

The effect of the selection on the inclusive sample of $\ttbar$ events is summarized in Table~\ref{tab:kineff}. The relative contribution from the leptonic channels after all the cuts is small, about 4\%. The distribution of the data events as a function of jet multiplicity is shown in Table~\ref{tab:jetmult}. Note that the requirement of the multijet trigger coupled with those of the kinematic selection modify the monotonically falling multiplicity spectrum of QCD background production processes.  

%%%%%%%%%%%%%%%%
\begin{table}[hbtp]
\begin{center}
\caption{Acceptance of the kinematic selection measured from {\sc PYTHIA} $\ttbar$ Monte Carlo simulation for $m_{\rm top}$=178\,GeV/$c^2$ and number of events selected in data.}\label{tab:kineff}
\begin{tabular}{lrr}
\hline\hline
Quantity & Acceptance (\%) & Data \\
\hline
Trigger & 63 & 4249644\\
Pre-selection & 42 & 3845744\\
$6\le N_{\rm jets}\le 8$ & 23 & 364006\\
${\cal{A}} + 0.005 \sum _3\Et\ge 0.96$ & 9.5 & 9425 \\
$C\ge 0.78$ & 6.9 & 3880\\
$\sum \Et\ge 280$\,GeV & 6.7 & 3315\\
\hline\hline
\end{tabular}
\end{center}
\end{table}

%%%%%%%%%%%
\begin{table}[hbtp]
\begin{center}
\caption{The number of events before and after kinematic selection for different jet multiplicities}
\label{tab:jetmult}
\begin{tabular}{c c c}
\hline\hline
Jet & Events before & Events after \\
Multiplicity & kinematic selection & kinematic selection \\
\hline
4 & 1540858 & 101\\
5 & 854266  & 695 \\
6 & 282521  & 1369 \\
7 & 68317   & 1300 \\
8 & 13168   & 646 \\
\hline\hline
\end{tabular}
\end{center}
\end{table}
%%%%%%%%%

The systematic uncertainties affecting the $\ttbar$ acceptance are summarized in Table~\ref{tab:kineffsyst}. 
The systematic uncertainty of 19.4\% arising from the jet energy scale is dominant, since this analysis requires the presence of a large number of jets in the event which are used to build the set of kinematic variables employed in the selection. 
We also study the effects on the efficiency of different Monte Carlo physics generation schemes, initial and final state radiation ISR/FSR, and the variation of parton distribution functions PDFs within their uncertainties. 

%%%%%%%%%%%%%%%%
\begin{table}[hbtp]
\begin{center}
\caption{Relative systematic uncertainties on the signal acceptance.}\label{tab:kineffsyst}
\begin{tabular}{lr}
\hline\hline
Quantity & Relative error (\%) \\
\hline
Energy Scale & 19.4 \\
PDF          & 2.6 \\
ISR/FSR      & 4.2 \\ 
Monte Carlo Modeling & 1.7 \\
\hline
Total        & 20.1 \\
\hline\hline
\end{tabular}
\end{center}
\end{table}
%%%%%%%%%%%%%%%%

\section{\label{sec:btag}$b$-Tagging in the Multijet Sample}

In order to further improve the signal-to-background ratio, we exploit the heavy flavor content of $\ttbar$ events using a $b$-tagging algorithm based on secondary vertex reconstruction 
as described in detail in\,\cite{ljetsPRD,massTMT}. The algorithm identifies jets containing a $b$-hadron state by reconstructing its decay vertex with at least two good quality tracks with hits in the silicon vertex  detector. A $b$-tagged jet must have an associated secondary vertex with a displacement from the primary vertex in the transverse plane with a significance larger than 7.5, where the typical resolution of the vertex displacement is about 190\,$\mu$m.
The efficiency to tag real $b$ quarks and the average number of tags per event, $n^{\rm ave}_{\rm tag}$, the quantities used in the cross section calculation, are measured in $\ttbar$ Monte Carlo events after the complete kinematic selection. The method we use takes into account the different tagging efficiencies for jets coming from the fragmentation of  $b$-, $c$-, or light-flavored quarks. The rates for all possible combinations of heavy- and light-flavored jets in the events are measured and used to properly combine the different efficiencies. 
This is particularly important in the case of all-hadronic $\ttbar$ decays since we find that about 44\% of the events after the kinematic seclection contain a charm quark from a $W$ boson in a top decay and 17\% of the events contain two charmed quarks. In Table~\ref{tab:hf} we summarize the heavy flavor fractions in $\ttbar$ Monte Carlo events after kinematic selection. 
The efficiency calculation includes the correction factors $0.91\pm 0.06$ for $b$ jets and $0.91\pm 0.12$ for $c$ jets respectively. These factors account for the different efficiency measured in data and Monte Carlo events; their measurement is described in detail in\,\cite{ljetsPRD}.

We find that the average number of tags present in a $\ttbar$ event after kinematic selection is $n^{\rm ave}_{\rm tag}=0.846\pm 0.073$. The systematic uncertainty is dominated by the uncertainty of the data to Monte Carlo correction factors for tagging $b$ and $c$ jets, where 
the uncertainties on both factors are considered fully correlated. 

\begin{table}[hbtp]
\begin{center}
%%%uncomment for PRD
\squeezetable
\caption{Heavy flavor fractions (\%) for $\ttbar$ Monte Carlo events after kinematic selection.}\label{tab:hf}
\begin{tabular}{l c c c}
\hline\hline
\backslashbox{Number \\ of $b$ jets}{Number \\ of $c$ jets} & 0 & 1 & 2 \\
\hline
0 & $0.11\pm 0.01$ & $0.18\pm 0.02$ & $0.08\pm 0.01$\\
1 & $3.62\pm 0.07$ & $5.95\pm 0.09$ & $2.37\pm 0.06$\\
2 & $27.53\pm 0.17$ & $43.37\pm 0.19$ & $16.80\pm 0.14$ \\
\hline\hline
\end{tabular}
\end{center}
\end{table}

\section{\label{sec:bkg}Background Estimate}
The background sources for this final state are due mainly to QCD production of heavy-quark pairs ($b\bar b$ and $c\bar c$) and false tags from light-quark jets. Other standard model processes such as $W/Z$+jets can be neglected due to the smaller production cross section and small acceptance due to the selection cuts. 

Given the theoretical uncertainties related to the production cross section for the generation of N-parton events, it is important to have a method for the background estimate that does not require any Monte Carlo information, and thus, is based solely on data. The method we use is based on the fact that even if the relative contribution from different processes changes as a function of jet multiplicity, the probability that a fiducial jet, a jet with two good quality tracks in the silicon detector, is tagged is approximately constant with increasing multiplicity. 
This assumption allows us to use the tag rate extracted from events depleted in $\ttbar$ signal as a measure of the tag rate in events with higher jet multiplicity. The depleted events are taken to be those with exactly four jets.

\begin{figure}[htbp]
%\begin{center}
\centering
\includegraphics[width=5.0cm]{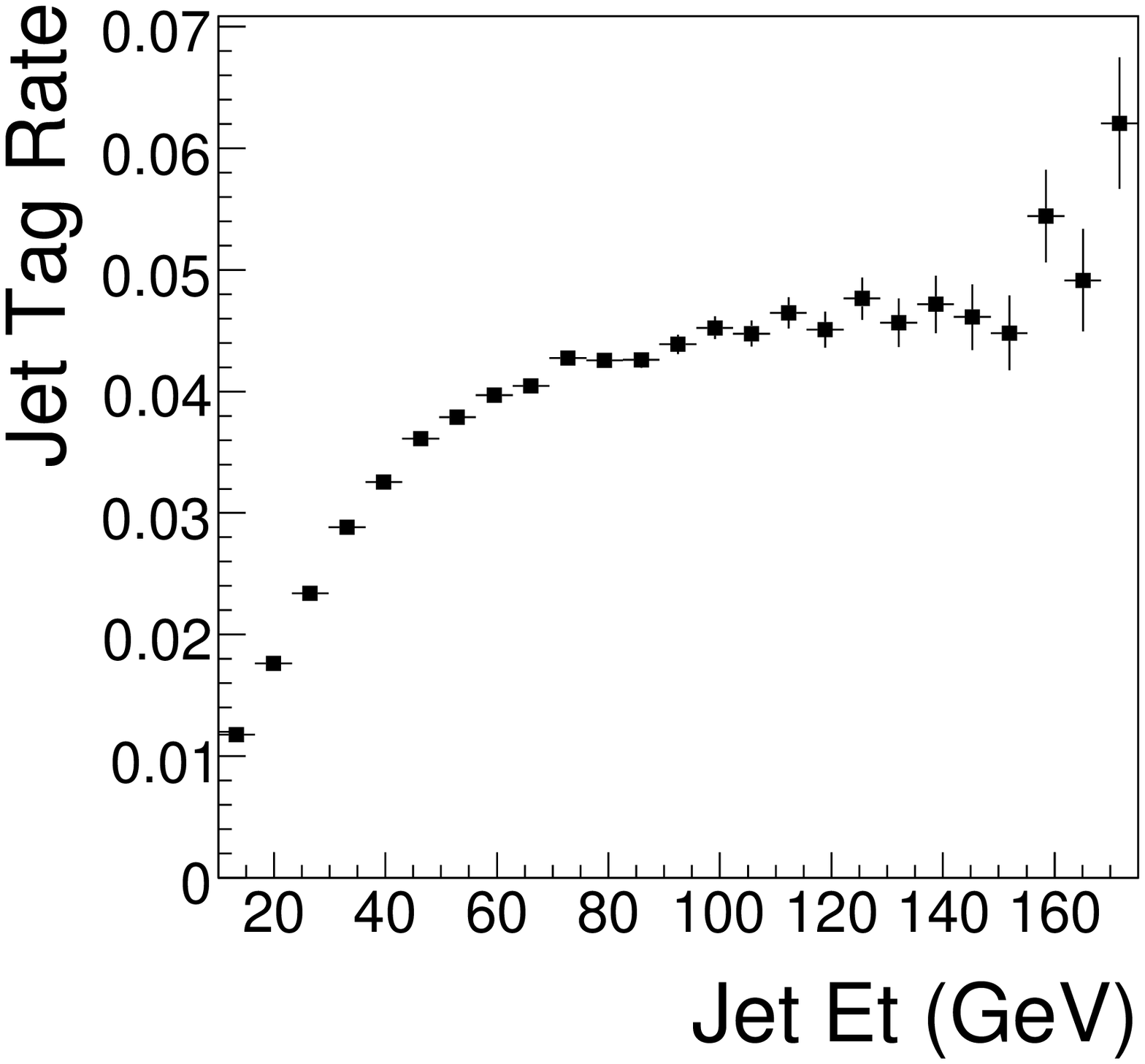}
\hfill
\includegraphics[width=5.0cm]{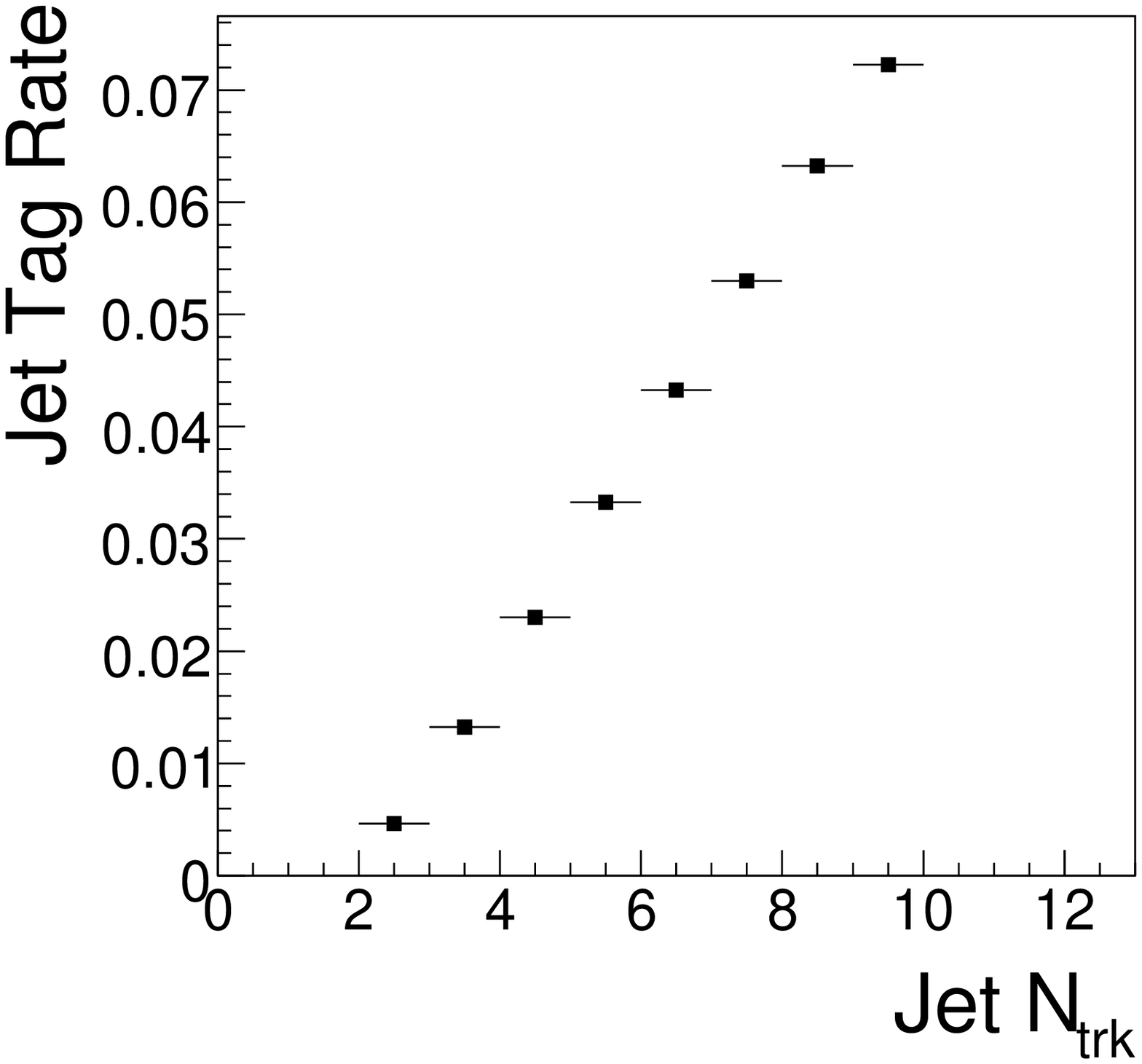}
\hfill
\includegraphics[width=5.0cm]{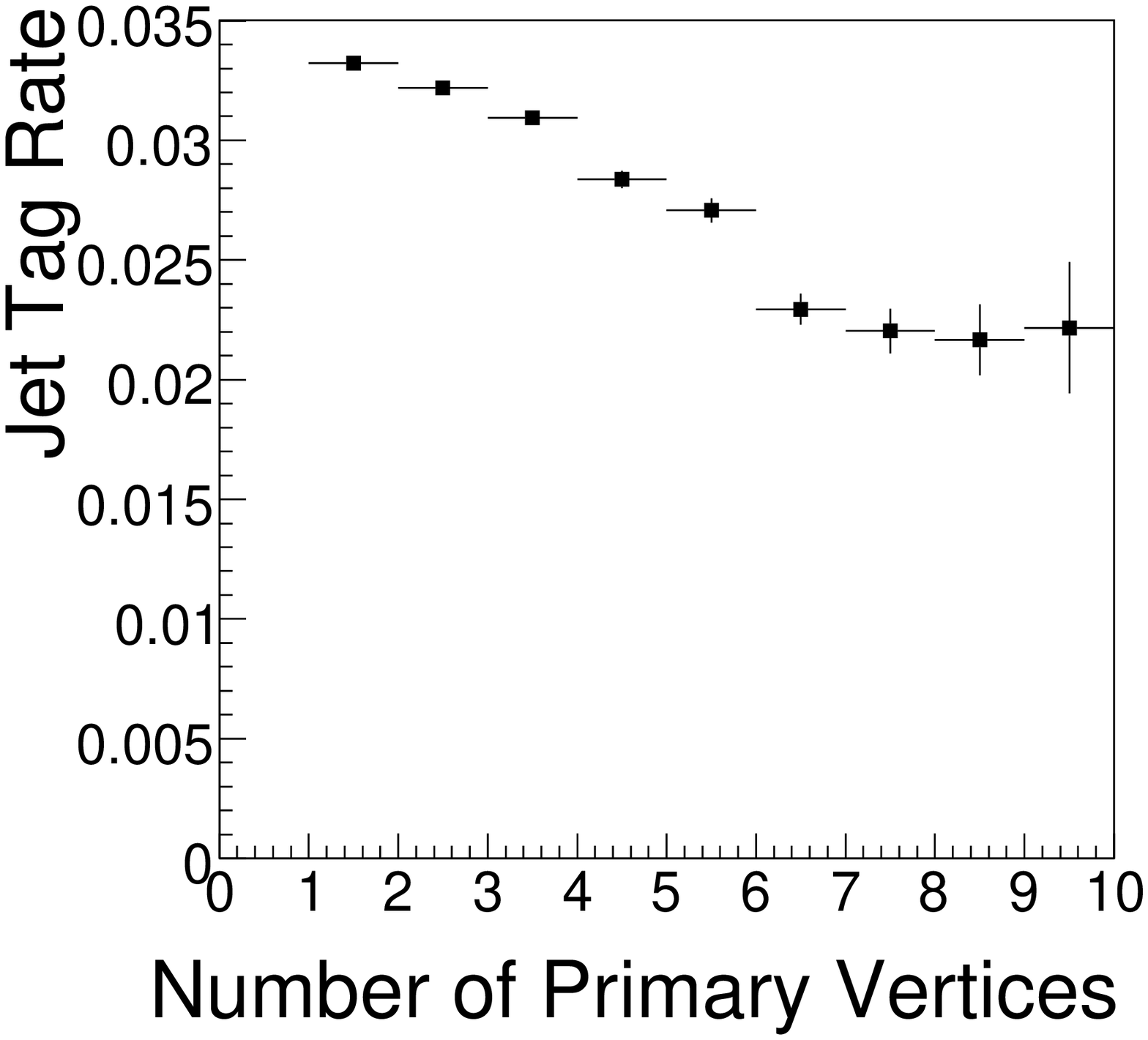}
\hfill
\caption{Tag rate for fiducial jets as a function of jet $\Et$, jet $N_{\rm trk}$, and $N_{\rm vert}$.} \label{fig:tagrate}
%\end{center}
\end{figure}

The tag rate per jet is evaluated in this $N_{\rm jets}=4$ sample and is parameterized in terms of variables sensitive to both the efficiency for true heavy-flavored objects and the rate of false tags.
These variables are jet-$\Et$, the number of tracks reconstructed in the vertex detector associated to the jet, $N_{\rm trk}$, and the number of primary vertices in the event, $N_{\rm vert}$. The tag rates per jet as a function of these variables are shown in Fig.~\ref{fig:tagrate}.

%%%%%%%%%%%%%%%%%
%%%%%%%%%%%%%%%%%%

\begin{figure}[htbp]
\begin{center}
\includegraphics[width=7.0cm]{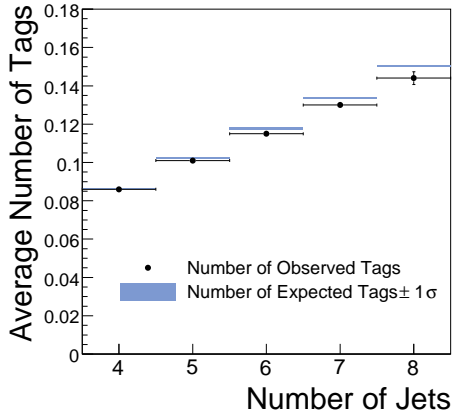}
\caption{Average number of tags per event observed in the multijet sample before kinematic selection compared with the estimate from the tag rate parametrization.} \label{fig:avebkg}
\end{center}
\end{figure}

This tag rate matrix provides the probability that a given fiducial jet in the signal sample is tagged. 
The expected number of tags from non-signal processes, that is QCD heavy and light-flavored production, in a set of selected events, is obtained by summing this jet tag probability over all fiducial jets in all the events.
Before the kinematic selection, the multijet sample is composed essentially of background events. The goodness of the parameterization and the goodness of the resulting estimate for different jet multiplicities is shown in Fig.~\ref{fig:avebkg}. 
The remaining small discrepancy of $2.1$\% observed at high jet multiplicity is accounted for as a systematic uncertainty on the background estimate. 

%%%%%%%%%%%%%%%%%
%%%%%%%%%%%%%%%%%%

The kinematic selection also changes the event characteristics with respect to those found in
 the 
sample with exactly four jets, where the parametrization has been derived. This selection modifies the jet-$E_T$ and $\eta$ spectra so that the average tag rate per event for jets from QCD background becomes higher. However, the parametrization of the tag probability in terms of properties of the jet ($E_T$, and $N_{\rm trk}$) is shown to describe well this increase for the kinematic selection.  
Possible biases due to the selection are treated as systematic uncertainties on the background prediction with the help of different control samples depleted in signal contribution. 
A specific control sample is defined for extracting the systematic uncertainty on the background determination due to each kinematic variable. The control sample is obtained by applying the $(N-1)$ kinematic selection cuts, and reversing the selection requirement on the chosen variable under study. For instance, in the case of the systematic error due to the $\sum E_T$ requirement, we apply the standard cuts on all other variables, and additionally require $\sum E_T\le 280$\,GeV. With this method we measure a relative systematic uncertainty of $4.1\%$ on the background estimate due to the kinematic selection by summing in quadrature the uncertainties obtained separately for each kinematic variable. 
The contributions from running conditions, such as instantaneous luminosity and detector configuration, have been studied and found to be negligible. 
After the application of the kinematic selection to a multijet sample of 311\,pb$^{-1}$ we are left with 3315 events with $6\le N_{\rm jets}\le 8$ in which there are 816 tags in 695 events. The distribution of observed tags and events for the different jet multiplicities is shown in Table~\ref{tab:tagskin}.

After kinematic selection, the expected background is 
$717\pm 29$ tags. However, since this background estimate is obtained from all the events passing the selection before tagging we need to subtract the contribution due to the $t\bar t$ events. This amount is derived with an iterative procedure using the $t\bar t$ cross section from the data, that is the observed excess divided by the average number of tags.
After this correction, the number of tags expected from background sources is reduced to to $683.7\pm 37.5$ tags.
%%%%%%%%%%%%%%%%%
\begin{figure}[htbp]
\begin{center}
\includegraphics[width=7cm]{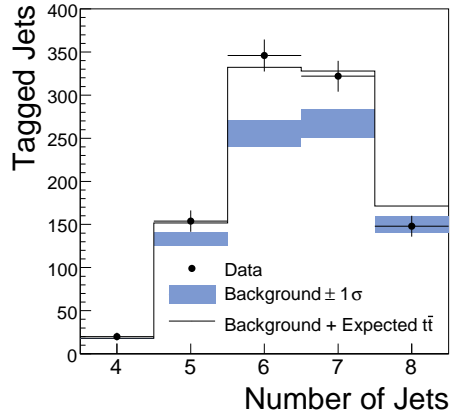}
\caption{Number of tags observed in multijet data after kinematic selection compared with the expected background. The $\ttbar$ expectation refers to the measured cross section of 7.5\,pb.} \label{fig:tagskin}
\end{center}
\end{figure}
%%%%%%%%%%%%%%%%%%

\begin{table*}[hbtp]
\begin{center}
\caption{Observed number of tags and expected background and signal after the kinematic selection. Only one cumulative value is given for the corrected background in $6\le N_{jet}\le 8$ because the iterative correction is applied to all the entries in the signal region.}\label{tab:tagskin}
\begin{tabular}{c cc|ccc}
\hline\hline
Jet Multiplicity & 4 & 5 & 6 & 7 & 8 \\
\hline
Background & $18.27\pm 0.55$ & $139.6\pm 5.8$& $283.5\pm 11.7$& $284.9\pm 11.7$ & $148.8\pm 6.1$ \\
Corrected Background & $17.6\pm 0.3$ & $133.4\pm 7.8$& \multicolumn{3}{c}{$683.7\pm 37.5$} \\
$\ttbar$ ($\sigma=6.1$~pb) & $0.5\pm 0.1$ & $14.7\pm 3.2$& $52.9\pm 11.6$& $39.3\pm 8.6$&$13.8\pm 3.0$%%% uncertainty on tt 20%+9%= 22% 
\\
\hline
Data & 20 & 154 & 346 & 322 & 148\\
\hline\hline
\end{tabular}
\end{center}
\end{table*}
%%%%%%%%%%%%%%%%

\section{\label{sec:xs}Cross Section Measurement}
The excess of the observed data over the background in the signal region shown in Table~\ref{tab:tagskin} is ascribed to $\ttbar$ production. A measurement of the cross section can be extracted from the acceptance and the background estimate: 

$$\sigma_{\ttbar}=\frac{N_{\rm obs}-N_{\rm bkg}}{\epsilon_{\rm kin} \times n^{\rm ave}_{\rm tag}\times {\cal{L}}_{\rm int}},$$
where $N_{\rm obs}=816$ and $N_{\rm bkg}=684\pm 38$ are the number of total observed and background tags, respectively, in the signal region $6\le N_{\rm jets}\le 8$, $\epsilon_{\rm kin}=6.7\pm 1.4\%$ is the signal kinematic selection efficiency shown in Table~\ref{tab:kineff}, $n^{\rm ave}_{\rm tag}=0.846\pm 0.073$ is the average number of tags in $\ttbar$ events and ${\cal L}_{\rm int}=311\pm 18$\,pb$^{-1}$ is the integrated luminosity. 
The value of the $\ttbar$ cross section is: 
$\sigma_{\ttbar}=7.5\pm 2.1({\rm stat.})^{+3.3}_{-2.2}({\rm syst.})^{+0.5}_{-0.4}({\rm lumi.})$\,pb for a top mass of 178\,GeV/$c^2$.
In Fig.~\ref{fig:tagskin} the distribution of the number of observed tags and background is compared to the $\ttbar$ signal expectation assuming the production cross section measured in this analysis. 

The cross section is also measured for different top quark mass assumptions as shown in Table~\ref{tab:xsmass}. In the same table are reported the kinematic efficiency and the relative systematic uncertainty due to the jet energy scale. The dependence of the average number of tags on $m_{\rm top}$ has been found to be negligible.

%%%%%%%%%%%%%%%%
\begin{table}[hbtp]
\begin{center}
\caption{Kinematic selection efficiency, relative systematic uncertainty from jet energy scale (JES), and measured cross section for different top quark mass assumptions.}\label{tab:xsmass}
\begin{tabular}{lccr}
\hline\hline
$m_{\rm top}$(GeV/$c^2$) & $\epsilon_{\rm kin}(\%)$ & JES Syst.(\%) & $\sigma$(pb)\\
\hline
165 & $5.1\pm 1.1$ & 22.1 & $9.9^{+5.1}_{-4.0}$\\
170 & $5.8\pm 1.1$ & 21.2 & $8.7^{+4.4}_{-3.4}$\\
175 & $6.3\pm 1.3$ & 20.1 & $8.0^{+4.2}_{-3.2}$\\
180 & $6.9\pm 1.3$ & 18.8 & $7.3^{+3.7}_{-3.0}$\\
185 & $7.6\pm 1.3$ & 17.1 & $6.6^{+3.3}_{-2.7}$\\
\hline\hline
\end{tabular}
\end{center}
\end{table}
%%%%%%%%%%%%%%%%

\section{\label{sec:fine}Conclusions}

Using an optimized kinematic selection and a $b$-jet-identification technique, we are able to improve the S/B of the initial multijet sample obtained with a dedicated trigger from 1/3500 to 1/5. 
With the selected sample, we measure the production cross section for $\ttbar$ events in the all-hadronic final state to be 
$\sigma_{\ttbar}=7.5\pm 2.1({\rm stat.})^{+3.3}_{-2.2}({\rm syst.})^{+0.5}_{-0.4}({\rm lumi.})$\,pb 
assuming $m_{{\rm top}}=178$\,GeV/$c^2$. 
These results agree well with the standard model expectation of $\sigma_{\ttbar}=6.1$\,pb for the same value of the top mass and with the results obtained in the leptonic channels. 
The current all-hadronic measurement is dominated by systematic uncertainties. The increase in integrated luminosity we expect from Run II will not only reduce the statistical uncertainty but will also allow for a more stringent selection with a better signal-to-background ratio. 
In particular the application of strategies based on neural network selection and the requirement of two identified $b$-quark jets per event can help to achieve a signal-to-background ratio of about 1/1 and a significant reduction in the systematic uncertainty. 

\begin{acknowledgments}
%%\section{Acknowledgments}

We thank the Fermilab staff and the technical staffs of the participating institutions for their vital contributions. This work was supported by the U.S. Department of Energy and National Science Foundation; the Italian Istituto Nazionale di Fisica Nucleare; the Ministry of Education, Culture, Sports, Science and Technology of Japan; the Natural Sciences and Engineering Research Council of Canada; the National Science Council of the Republic of China; the Swiss National Science Foundation; the A.P. Sloan Foundation; the Bundesministerium f\"ur Bildung und Forschung, Germany; the Korean Science and Engineering Foundation and the Korean Research Foundation; the Particle Physics and Astronomy Research Council and the Royal Society, UK; the Russian Foundation for Basic Research; the Comisi\'on Interministerial de Ciencia y Tecnolog\'{\i}a, Spain; in part by the European Community's Human Potential Programme under contract HPRN-CT-2002-00292; and the Academy of Finland. 

\end{acknowledgments}


\begin{thebibliography}{99}
%9 

%00
%01
\bibitem{topxs} M. Cacciari {\it et al.}, JHEP 0404:068 (2004); N. Kidonakis and R. Vogt, Phys. Rev. {\bf D68}, 114014 (2003)
%2
\bibitem{RunIavemass} CDF Collaboration, D\O\ Collaboration, and the Tevatron Electroweak Working Group, hep-ex/0404010 (2004).
%3
\bibitem{tophadPRL} F.~Abe {\it et al.} (CDF Collaboration), Phys. Rev. Lett. {\bf 79}, 1992 (1997); B.~Abbott {\it et al.} (D\O\,Collaboration), {\it ibid} {\bf 83}, 1908 (1999). 
%4
\bibitem{dilPRL} F.~Abe {\it et al.} (CDF Collaboration), Phys. Rev. Lett. {\bf 93}, 142001 (2004).
%5
\bibitem{ljetsPRD} D.~Acosta {\it et al.} (CDF Collaboration), Phys. Rev. {\bf D71}, 052003 (2005).
%6
\bibitem{ljetsKIN} D.~Acosta {\it et al.} (CDF Collaboration), Phys. Rev. {\bf D72}, 052003 (2005).
%7
\bibitem{ljetsBKIN} D.~Acosta {\it et al.} (CDF Collaboration), Phys. Rev. {\bf D71}, 072005 (2005).
%8
\bibitem{ljetsSLT}D.~Acosta {\it et al.} (CDF Collaboration), Phys. Rev. {\bf D72}, 032002 (2005).

%9
\bibitem{CDFdetector} D.~Acosta {\it et al.} (CDF Collaboration), Phys. Rev. {\bf D71}, 032001 (2005).

%10
\bibitem{coordinate} We use a cylindrical coordinate system where $\theta$ is the polar angle to the proton beam direction at the event vertex, $\phi$ is the azimuthal angle about the beam axis, and pseudorapidity is defined $\eta = - \ln \tan(\theta/2)$. We define transverse energy as $\Et = E \sin\theta$ and transverse momentum as $p_T = p\sin\theta$ where $E$ is the energy measured in the calorimeter and $p$ is the magnitude of the momentum measured by the spectrometer.

%11 
\bibitem{ecal} L.~Balka {\it et al.}, Nucl. Instrum. Methods Phys. Res., Sect A {\bf A67}, 272 (1988).
%12
\bibitem{pem} M.~Albrow {\it et al.}, Nucl. Instrum. Methods Phys. Res., Sect A {\bf 480}, 524 (2002).
%13 TDR replaces the reference above for the calorimeter resolution
\bibitem{TDR} R.~Blair {\it et al.}, {\it The CDF-II detector: Technical Design Report}, FERMILAB-PUB-96-390-E.
%14
\bibitem{cha} S.~Bertolucci {\it et al.}, Nucl. Instrum. Methods Phys. Res., Sect A {\bf 267}, 301 (1988).
%%14 ``topology of 3 jets events'' to reference the jet clustering algorithm
\bibitem{jets} F.~Abe {\it et al.} (CDF Collaboration), Phys. Rev. {\bf D45}, 1448 (1992).
%14bis CMU reference
\bibitem{CMU} G.~Ascoli {\it et al.}, Nucl. Instrum. Methods Phys. Res., Sect A {\bf 268}, 33 (1988).
%15 lummi
\bibitem{WZprl} F.~Abe {\it el al.} (CDF Collaboration), Phys. Rev. Lett. {\bf 94}, 091803 (2005).
%16
\bibitem{JESNIM} A.~Bhatti {\it et al.}, hep-ex/0510047 (2005).
%17
\bibitem{metjetPRL} A.~Abulencia {\it et al.} (CDF Collaboration), Phys. Rev. Lett {\bf96}, 202002 (2006). 

\bibitem{met} The missing transverse energy $\vec{\met}$ is calculated as the vector sum of the energy in each calorimeter tower multiplied by a unit vector in the azimuthal direction of the tower. If isolated high momentum muons are found in the event, the $\vec{\met}$ is corrected by subtracting the muon energy in the calorimeter and adding the muon $p_T$ to the vector sum.  $\met$ is defined as the magnitude of  $\vec{\met}$.

%18

%19
\bibitem{Pythia} T.~Sjostrand {\it et al.}, Comput. Phys. Commun. {\bf 135}, 238 (2001).
%20
\bibitem{Herwig} G.~Marchesini {\it et al.}, Comput. Phys. Commun. {\bf 67}, 465 (1992); G.~Corcella {\it et al.}, J. High Energy. Phys. {\bf 0101}, 010 (2001).
%21
\bibitem{massTMT} A.~Abulencia {\it et al.} (CDF Collaboration), Phys. Rev. {\bf D73}, 032003 (2006).

\end{thebibliography}
\end{document}